\documentclass[aps,prb,superscriptaddress,twocolumn]{revtex4-2}
\usepackage{nicefrac}
\usepackage{graphicx}
\usepackage{dcolumn}
\usepackage{epstopdf}
\usepackage{bm}

\usepackage{braket}

\usepackage[pdftex,
colorlinks=true,
citecolor=blue,
setpagesize=false]{hyperref}

\begin{document}

\preprint{APS/123-QED}

\title{Direction-selective intertwined charge, orbital, and lattice orders under uniaxial strain in hole-doped manganite: La$_{0.75}$Ca$_{0.25}$MnO$_3$}%

\author{Ju Hyeon Lee}
\affiliation{Department of Physics, Kyungpook National University, Daegu 45166, Republic of Korea}
\author{Beom Hyun Kim}
\email{bomisu@gmail.com}
\affiliation{Department of Physics \& Astronomy, Seoul National University, Seoul 08826, Republic of Korea}
\author{Bongjae Kim}
\email{bongjae@knu.ac.kr}
\affiliation{Department of Physics, Kyungpook National University, Daegu 45166, Republic of Korea}

\date{\today} 

\begin{abstract}
 The complex interplay of charge, spin, orbital, and lattice degrees of freedom governs emergent phases in quantum materials, making strain a powerful control parameter. Recent advances in free-standing layer techniques have enabled extreme strains of nearly 8\%, opening access to novel and often unexpected electronic and magnetic phases. Here, using a density functional theory approach, we investigate the effect of direction-selective uniaxial strain on the prototypical Jahn-Teller system La$_{1-x}$Ca$_x$MnO$_3$ ($x=0.25$). We find that different strain directions stabilize qualitatively distinct structural, charge, and orbital responses, rather than merely different strengths of the same phase. In particular, extreme uniaxial strain selectively induces cooperative Jahn-Teller, breathing-like, and site-selective modulations, thereby enabling previously inaccessible intertwined orders in manganites. These results establish direction-selective uniaxial strain as a powerful and selective route for engineering emergent phases in quantum materials.
\end{abstract}

\maketitle

\section{Introduction}

 Transition metal oxides (TMOs) provide a prototypical platform for emergent phenomena such as multiferroicity, superconductivity, and metal-insulator transitions (MIT), owing to the strong coupling among charge, spin, orbital, and lattice degrees of freedom~\cite{Imada1998,Rao1989}. While this complexity makes their microscopic behavior challenging to describe, it also offers broad opportunities for material design. External perturbations such as chemical doping, epitaxial strain, and heterostructure engineering can be used to tune these coupled degrees of freedom and thereby access new electronic and magnetic phases~\cite{Ramirez1997,Schlom2007,Ngai2014,Chen2019,Li2024_2,Goodenough2014,Okamoto2025}.

 LaMnO$_3$ (LMO) is a prototypical system of such intertwined physics. In the parent compound, the cooperative Jahn-Teller (JT) distortion, orbital order, and $A$-type antiferromagnetic (AFM) spin order are closely linked~\cite{Goodenough1955,He2012,Park1996,Imada1998,Pavarini2010,Lee2025,Dagotto2001,Yamasaki2006,Munkhbaatar2018,Celiberti2026}, and an explanation based on the Kugel-Khomskii physics has provided the coherent picture of the intertwined orders~\cite{Khomskii2021,Pavarini2010}. Upon hole doping by substitution of La$^{3+}$ with Ca$^{2+}$ or Sr$^{2+}$, additional carriers weaken the orbital order, promote metallicity, and alter the delicate balance between ferromagnetic (FM) and AFM exchange interactions~\cite{Sagdeo2006,Kliment1982,Mryasov1997,Mizokawa1999,Hwang1995,Ramirez1997,Hotta2000,Hotta2001,Aliaga2003,Dong2012,Radaelli1997}. At higher doping, as in La$_{0.5}$Ca$_{0.5}$MnO$_3$, coupled charge, orbital, and spin orders so-called the $CE$-type phase emerge~\cite{Hotta2000,Hotta2001,Aliaga2003,Dong2012,Radaelli1997}, highlighting the strong tunability of the intertwined orders in this family. This is schematically shown in Fig.~\ref{fig1}.

 Strain engineering offers another powerful route to controlling these competing phases~\cite{Robert1992,Li2024_2,Rondinelli2011}. Conventional substrate-based epitaxy has already revealed strain-induced MIT and magnetic phase transition in manganites~\cite{Ogimoto2005,Roqueta2015,Renshaw2015,Banerjee2020,Zhang2022,Guo2024,Li2024,Adamo2009,Vailionis2011,Zhang2021,Takamura2008,Wu1999}. However, the accessible strain range is often limited by substrate constraints and lattice relaxation. Recent advances in free-standing membrane techniques have substantially expanded this range, enabling extreme strains approaching {8\% in uniaxial and 5\% in biaxial geometries~\cite{Lu2016,Zhang2024}. In La$_{0.7}$Ca$_{0.3}$MnO$_3$, such extreme strain has already been shown to induce phases beyond those accessible in conventional regimes~\cite{Hong2020}.
 
 These developments raise an important question: can uniaxial strain act not only as a stronger perturbation, but also as a qualitatively new tuning knob for intertwined orders? In an orthorhombic manganite, different strain directions need not produce the same response with different magnitudes; rather, they may selectively couple to octahedral tilting, rotation, and JT distortions in distinct ways, thereby stabilizing different charge, orbital, and magnetic states.
 
 In this study, partly motivated by a previous investigation into extreme tensile strain~\cite{Hong2020}, we investigate how extreme, direction-selective uniaxial strain reshapes the intertwined orders in hole-doped La$_{0.75}$Ca$_{0.25}$MnO$_3$ employing density functional theory (DFT) based analysis. We systematically apply uniaxial strain along the (100), (010), (110), and ($\bar1\bar10$) directions of orthorhombic LCMO and examine the resulting structural, orbital, charge, and magnetic responses. We show that different strain directions do not merely tune the strength of the same phase, but instead stabilize qualitatively distinct intertwined orders.
In particular, strains along the (100) and (010) directions produce contrasting cooperative JT- and breathing-dominated responses because of the underlying orthorhombic anisotropy, whereas strains along the (110) and ($\bar1\bar10$) directions stabilize a new intertwined phase, which we denote as $CF$ order, characterized by the coexistence of ferro-orbital order and $C$-type charge disproportionation with a FM ground state. Our results establish direction-selective uniaxial strain as a powerful route to accessing previously unreachable intertwined orders in correlated oxides.

\begin{figure}
\centering
\includegraphics[width=\linewidth,clip,trim=1 1 1 1]{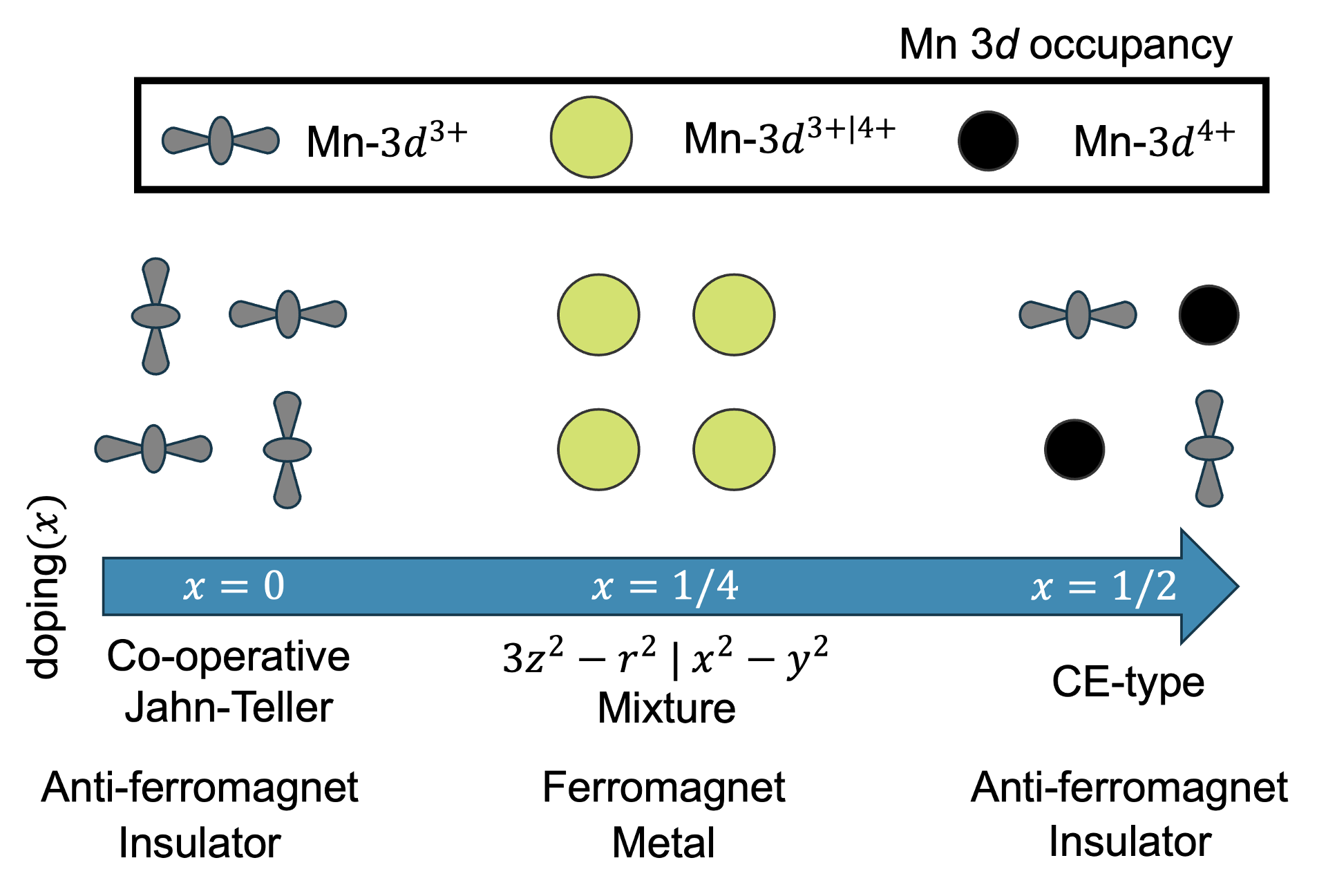}
\caption{\label{fig:left} Schematic phase diagram of bulk La$_{1-x}$Ca$_x$MnO$_3$ as a function of doping, $x$. At $x=0$, cooperative Jahn-Teller (JT) order, accompanied by $\ket{3x^2-r^2}$/$\ket{3y^2-r^2}$ orbital ordering, is stabilized within an $A$-type antiferromagnetic (AFM) insulating phase. At $x=1/4$, this orbital order is suppressed and the ferromagnetic (FM) metallic phase becomes the ground state. At half doping ($x=1/2$), a $CE$-type orbital and antiferromagnetic ordered insulating phase emerges.}
\label{fig1}
\end{figure}

\begin{figure*}[t]
\centering
\includegraphics[width=\linewidth,clip,trim=1 1 1 1]{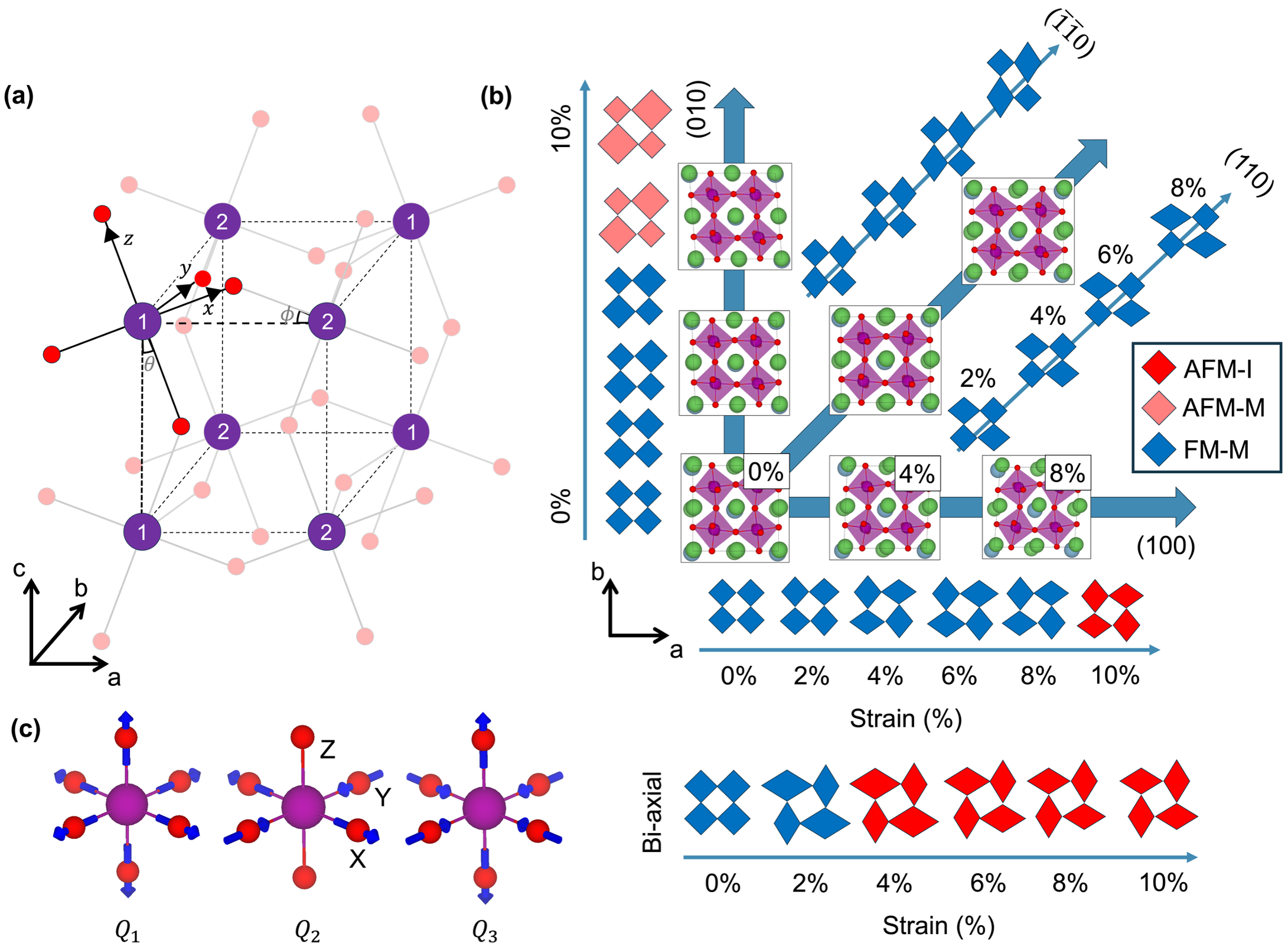}
\caption{\label{fig:left} (a) Schematic pseudocubic lattice of La$_{0.75}$Ca$_{0.25}$MnO$_3$ (LCMO), indicated by dashed lines. The two inequivalent Mn sites are labeled 1 and 2. $\phi$ and $\theta$ denote the in-plane rotation and out-of-plane tilting angles, respectively. Purple and red spheres represent Mn and O atoms. The $a$, $b$, and $c$ axes denote the global crystallographic axes, while $x$, $y$, and $z$ denote the local axes of the MnO$_6$ octahedron. (b) Schematic structural, electronic, and magnetic phase diagrams for different strain directions. Rhombus represent MnO$_6$ units, with their shapes and sizes indicating the type and magnitude of the distortion. Red, light red, and blue indicate $A$-type antiferromagnetic insulator (AFM-I), $A$-type antiferromagnetic metal (AFM-M), and ferromagnetic metal (FM-M) ground states, respectively. (c) Schematic illustration of the symmetric breathing mode $Q_1$ and the asymmetric Jahn-Teller modes $Q_2$ and $Q_3$, corresponding to orthorhombic and tetragonal distortions, respectively. Here, $X$, $Y$, and $Z$ are the Mn-O bond lengths along the local $x$, $y$, and $z$ axes.
\label{fig2}
}\end{figure*}

\section{Computational details}
 DFT calculations were performed using the Vienna \textit{ab initio} Simulation Package (VASP) with an energy cutoff of 600~eV and an energy convergence criterion of $10^{-9}$~eV~\cite{Kresse1996}. For the exchange-correlation functional, Perdew–Burke–Ernzerhof revised for solids (PBEsol) was employed~\cite{Perdew2008}. To account for the strong octahedral rotation and tilting of the MnO$_6$ units (see Fig.~\ref{fig2}(a)), we adopted an orthorhombic $Pnma$ unit cell containing 4 formula units (f.u.), corresponding to a $\sqrt{2}\times\sqrt{2}\times2$ supercell of cubic ABO$_3$. When uniaxial strain is applied, additional structural and charge orders are not captured within conventional 4 f.u. $Pnma$ cell. We therefore further expanded the cell by $\sqrt{2}\times\sqrt{2}$ in the $ab$ plane, resulting in an 8 f.u. $2\times2\times2$ supercell. To approximate the experimentally relevant doping level of $x\approx0.3$, we adopted a $x=0.25$ composition, corresponding to La$_{0.75}$Ca$_{0.25}$MnO$_3$. A Monkhorst-Pack $k$-point mesh of $4\times4\times4$ was used.

 To simulate strain, we fixed the in-plane lattice parameters ($a$ and $b$) to the corresponding strained values and fully relaxed all remaining structural degrees of freedom until the residual forces were smaller than $10^{-7}$~eV/\AA. To treat the electronic correlations in the localized Mn-$3d$ orbitals, we employed the Hubbard correction with an on-site Coulomb interaction $U=4$~eV and Hund's coupling $J_H=1$~eV~\cite{Park2015,Lee2025}. We note that the optimized structures show no significant dependence on the Hubbard parameters within the range considered. The chosen $U$ value also yields structural parameters in good agreement with the fully relaxed experimental crystal structure (see Supplementary Table~S.1). Our calculations for $x=0.25$ do not exhibit the strain-induced metal-insulator transition reported experimentally near $x=0.3$; this difference is discussed in detail in Sec.~III.D.

\section{Results and discussions}
\subsection{Phase diagram}
 In Fig.~\ref{fig2}(b), we schematically present the full phase diagram of uniaxially strained LCMO along four representative crystallographic directions: (100), (010), (110), and ($\bar1\bar10$). The relation between these axes and the pseudocubic LCMO unit cell is illustrated in Fig.~\ref{fig2}(a). (See also Supplementary Fig.~S1 for the geometric relation between the orthorhombic and pseudocubic coordinate systems used in this work.) Notably, the relevant phases show a clear dependence on the strain direction, which is also distinct from the biaxial-strain case.

 This directional selectivity originates from the orthorhombic $Pnma$ symmetry of LCMO, in which the MnO$_6$ octahedra exhibit the $a^-a^-c^+$ tilting pattern in Glazer notation. As a result, the (100) and (010) directions are not symmetrically equivalent. Under uniaxial strain along the (100) direction, the cooperative JT distortion progressively develops, whereas under strain along the (010) direction, a breathing-like modulation becomes more prominent. While the former is commonly found in manganites~\cite{Zhang2022}, the latter is unusual and more often associated with nickelates~\cite{Park2012,Bayo2013,Johnston2014,Fernandez2000,Medarde2009}. This contrast demonstrates that even within the same material, different uniaxial strain directions selectively promote distinct lattice responses, which are further reflected in the electronic structure and in the strain-driven transition from FM to an $A$-type AFM ground state (see Fig.~\ref{fig2}(b)).

 The response under diagonal strain, along the (110) and ($\bar1\bar10$) directions, is qualitatively different. Rather than being a simple combination of the (100) and (010) direction cases, they stabilize a distinct intertwined phase in which the two types of MnO$_6$ octahedra develop inequivalent structural and electronic characters, while the FM ground state is retained. We denote this phase as $CF$ order, characterized by the coexistence of ferro-orbital order in the same types of MnO$_6$ octahedra, $C$-type charge disproportionation in distinct types, and a FM metallic state. While this phase is reminiscent of the coupled charge-orbital textures of conventional $CE$-type ordering, it differs fundamentally in its metallic ferromagnetic nature and ferro-orbital order in the same types of MnO$_6$ octahedra. The emergence of this phase points to physics beyond Kugel-Khomskii picture~\cite{Khomskii2021}.

 We also compared our phase diagram with that obtained under extreme biaxial strain. The FM-to-AFM phase transition has been proposed as the origin of the MIT in this system~\cite{Hong2020}, and this behavior is reproduced in our calculations. Under biaxial strain, breathing distortion and cooperative JT modes evolve in a more intertwined manner, and the magnetic transition is triggered at lower strain than in the uniaxial cases (see Fig.~\ref{fig2}(b)). In our calculations, the transition occurs at approximately 4\% strain, slightly higher than the 3\% reported previously, likely reflecting differences in the computational setup. The magnetic transition data for all studied strain directions are shown in Supplementary Fig.~S2.

 Overall, the stark differences in the structural, electronic, and magnetic responses among the four uniaxial strain directions demonstrate that extreme uniaxial strain acts as a highly selective tuning knob that goes beyond the capabilities of conventional substrate-based biaxial strain engineering. In the following sections, we show how these distinct phases arise from the underlying local structural modulations.

\begin{figure}[t]
\centering
\includegraphics[width=\linewidth,clip,trim=1 1 1 1]{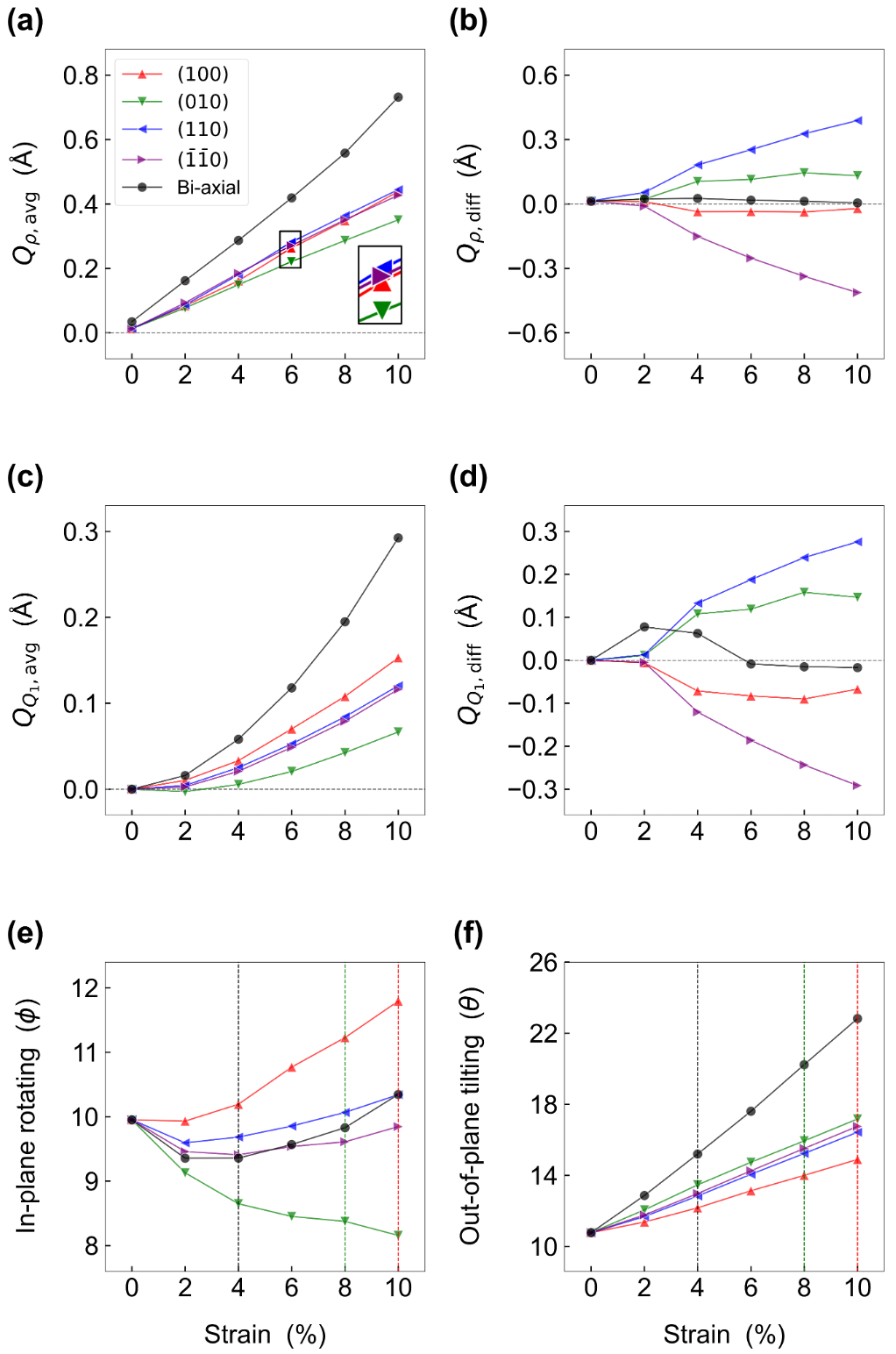}
\caption{\label{fig:left} Strain dependence of the averaged and site-difference of the asymmetric Jahn-Teller mode in (a) and (b), and of the breathing mode in (c) and (d), under uniaxial and biaxial strain. (e) and (f) show the in-plane rotation angle $\phi$ and out-of-plane tilting angle $\theta$ of the MnO$_6$ octahedra, respectively, as defined in Fig.~\ref{fig2}(a). The inset provides an enlarged view of the closely overlapping curves for the (100), (110), and ($\bar1\bar10$) strain directions. The vertical dashed lines indicate the critical strains for the magnetic phase transitions, with colors corresponding to those of the respective strain directions.}
\label{fig3}
\end{figure}

\subsection{Structural response}

 To quantify the intertwined structural, charge, and orbital orders, we identify the relevant lattice modes of the system, as schematically shown in Fig.~\ref{fig2}(c). The structural response of LCMO can be described in terms of the symmetric mode $Q_1$ and the asymmetric JT modes $Q_2$ and $Q_3$, as is typical for $Pnma$ TMO perovskites. The $Q_1$ mode corresponds to a breathing-type distortion, whereas $Q_2$ and $Q_3$ represent orthorhombic and tetragonal distortions, respectively. The mode amplitudes are defined as $Q_{1,i}=\frac{X_i+Y_i+Z_i}{\sqrt{3}}$, $Q_{2,i}=\frac{X_i-Y_i}{\sqrt{2}}$, and $Q_{3,i}=\frac{2Z_i-X_i-Y_i}{\sqrt{6}}$, where $i$ denotes the MnO$_6$ site index, and $X_i$, $Y_i$, and $Z_i$ denote Mn--O bond-length differences along the local $x$, $y$, and $z$ axes, respectively (see Fig.~\ref{fig2}(a)). Unlike the conventional cooperative JT distortion, the extreme-strain regime exhibits pronounced site selectivity. In other words, a given mode develops cooperatively on both sites 1 and 2, or becomes enhanced predominantly on only one of them. To quantify this behavior, we define $Q_{1,\mathrm{avg}}=\nicefrac{\sum_{i=1,2}Q_{1, i}}{2}$ and $Q_{1, \mathrm{diff}}=\sum_{i=1,2}(-1)^{i+1}Q_{1, i}$, where the former and the latter represent the average amplitude and site difference of the $Q_1$ mode, respectively. For the asymmetric JT modes, we similarly define $Q_{\rho,\mathrm{avg}}=\nicefrac{\sum_{i=1,2}\rho_i}{2}$ and $Q_{\rho, \mathrm{diff}}=\sum_{i=1,2}(-1)^{i+1}\rho_i$, where $\rho_i=\sqrt{Q_{2,i}^2+Q_{3,i}^2}$ characterizes the combined amplitude of the $Q_2$ and $Q_3$ distortions at each site.

 A notable common feature of LCMO under extreme strain is the emergence of the breathing mode $Q_1$, regardless of the strain direction. A charge-ordered pattern of this type has not been reported in La$_{1-x}$Ca$_x$MnO$_3$ ($x \leq 0.5$) under conventional epitaxy-based effective strains, which are typically much smaller due to lattice relaxations \cite{Ziese2003, Siwach2006}. Moreover, the resulting two-dimensional checkerboard-like $Q_1$ modulation is fundamentally different from the breathing distortions reported in nickelates~\cite{Park2012,Bayo2013,Johnston2014} and other TMO systems~\cite{Ye2021,Prodi2014,Yang2018}. The appearance of $Q_1$ modulation in doped manganites, therefore, expands the diversity of accessible structural and electronic phases.

 We first consider the biaxial-strain case. As shown in Fig.~\ref{fig3}(a) and (b), the average JT distortion $Q_{\rho,\mathrm{avg}}$ increases almost linearly with strain, whereas $Q_{\rho,\mathrm{diff}}$ remains almost muted, indicating little site differentiation up to 10\% strain. The behavior of the breathing distortion is markedly different. As shown in Fig.~\ref{fig3}(c) and (d), $Q_{1,\mathrm{avg}}$ increases sharply, especially at and above 4\% strain, indicating that the $Q_1$ modulation emerges only beyond the conventional substrate-engineering regime. A finite $Q_{1,\mathrm{diff}}$ appears around 2 to 4\%, showing that the breathing distortion initially develops a site-selective character together with the cooperative JT response. At higher strain, however, this site selectivity is suppressed again, as evidenced by the reduction of $Q_{1,\mathrm{diff}}$.

 For the uniaxial strain, the structural response becomes strongly direction dependent. In the pseudocubic unit cell, strain applied along the (100) and (010) directions acts along diagonal directions (See Fig.~\ref{fig2}(b)) and therefore directly modifies the in-plane rotation of the MnO$_6$ octahedra. Because these two directions are not symmetry-equivalent in orthorhombic LCMO, their structural responses differ substantially. As shown in Fig.~\ref{fig3}(e), the in-plane rotation increases under strain along (100), but decreases under strain along (010), directly demonstrating the anisotropic structural response of the system.

 This inequivalence is also evident in the responses of the asymmetric JT and breathing modes. Both strain directions enhance the amplitudes of these modes, as shown in Fig.~\ref{fig3}(a) and (c), with a stronger overall modulation for strain along (100), consistent with the increase in in-plane octahedral rotation shown in Fig.~\ref{fig3}(e). However, the site selectivity is more pronounced for strain along (010). In particular, $Q_{\rho,\mathrm{diff}}$ becomes activated above 4\% strain for the (010) case, whereas it is vanishingly small for the (100) case (Fig.~\ref{fig3}(b)). Likewise, a nonzero $Q_{1,\mathrm{diff}}$ appears in both cases above about 4\% strain, but its magnitude is clearly larger under strain along (010). These results indicate that strain along (100) mainly enhances cooperative JT distortion and lifts the $e_g$ orbital degeneracy through orthorhombic distortion, whereas strain along (010) promotes a more site-selective response in which the breathing component plays a more prominent role. This distinction becomes clearer in the next subsection through the orbital-resolved analysis.

 For uniaxial strain along the (110) and ($\bar1\bar10$) directions, the response is qualitatively different. In these cases, the strain is applied directly along the Mn--O bond directions in the pseudocubic cell, which strongly restricts the in-plane octahedral rotation as shown in Fig.~\ref{fig3}(e). As a result, the lifting of the $e_g$ degeneracy is governed primarily by the asymmetric responses of the JT.

The site selectivity becomes especially strong for these diagonal strain directions, as seen in Fig.~\ref{fig3}(b). The JT distortion develops predominantly on one type of MnO$_6$ octahedron, while it is suppressed on the other. This asymmetry is mirrored in the breathing $Q_1$ mode. Indeed, Fig.~\ref{fig3}(d) shows the opposite trends for the two sites. These results demonstrate that the direction of uniaxial strain can selectively activate or suppress specific lattice responses, thereby determining which structural modulations become dominant. Notably, the FM ground state remains stable throughout the entire strain range studied~\cite{Sen2020}.

 We further note that $Q_{\rho,\mathrm{avg}}$ is significantly larger under biaxial strain than under uniaxial strain. At the same nominal in-plane strain value, biaxial strain imposes a larger two-dimensional in-plane stress, which in turn promotes stronger out-of-plane tilting, as shown in Fig.~\ref{fig3}(a) and (f). More detailed structural information, including lattice parameters, volume changes, and bond-length evolution under strain, is provided in  Supplementary Figs.~S3 and S4.

 An important ingredient underlying the diversity of strain-dependent behavior is hole doping. In undoped LMO, which has a similar orthorhombic structure, site-selective modulation is strongly suppressed for all uniaxial strain directions. Although the orthorhombicity-driven responses of the JT and breathing modes are qualitatively similar, the additional holes in LCMO promote instability and activate site-selective modulations, going beyond the behavior found in LMO~\cite{Rivero2016_2,Lee2013}. Detailed results for LMO are provided in Supplementary Fig.~S5.
 
\begin{figure}[t]
\centering
\includegraphics[width=\linewidth,clip,trim=1 1 1 1]{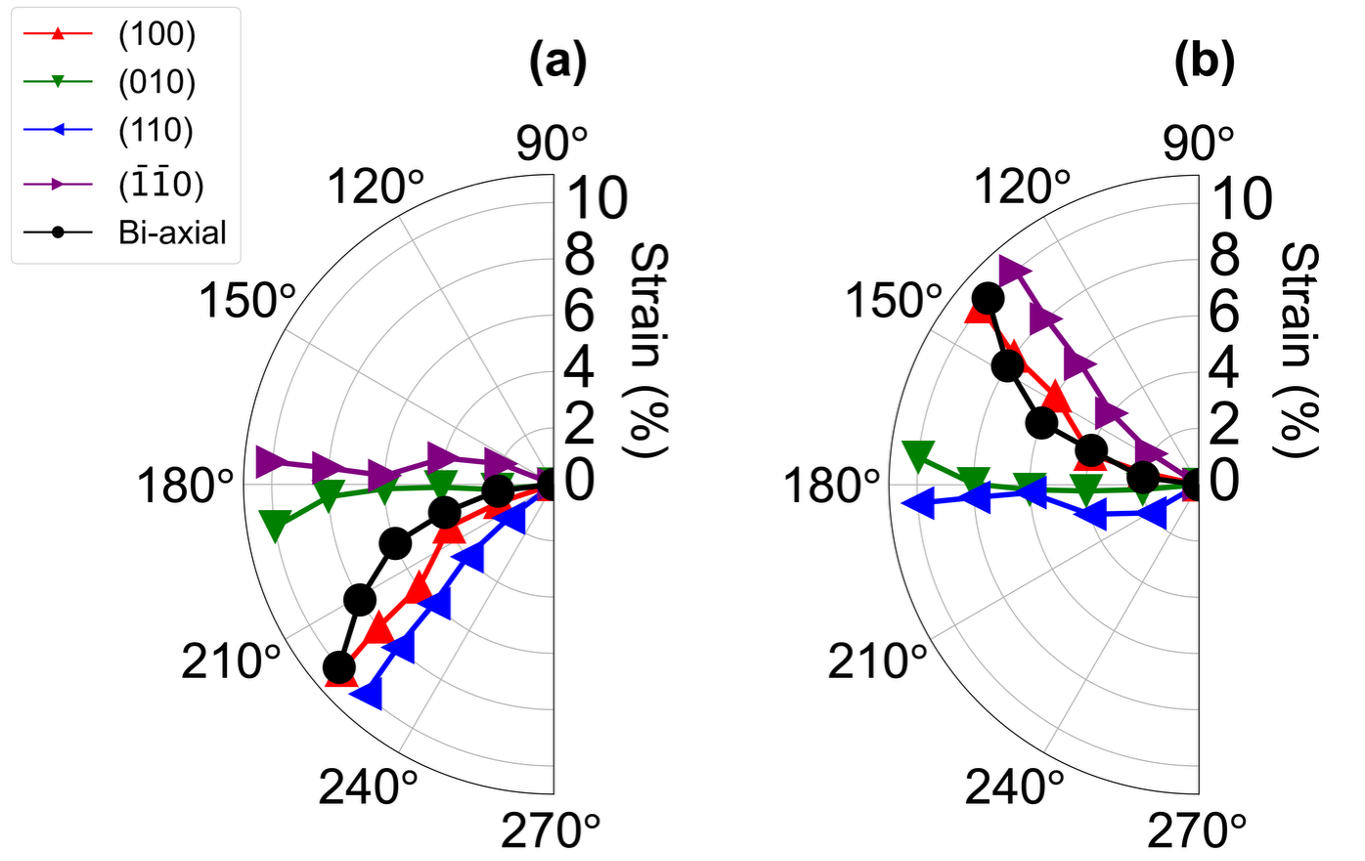}
\caption{\label{fig:left} Orbital angles for Mn sites 1 and 2, shown in (a) and (b), respectively, defined as $\tan^{-1}(Q_2/Q_3)$ (see Fig.~\ref{fig2}(a) for the site labels). The orbital characters $\ket{3x^2-r^2}$, $\ket{3y^2-r^2}$, and $\ket{x^2-y^2}$ correspond to orbital angles of $120^\circ$, $240^\circ$, and $180^\circ$, respectively. The black, red, green, blue, and purple curves represent biaxial strain and uniaxial strain applied along the (100), (010), (110), and ($\bar1\bar10$) directions, respectively.
}
\label{fig4}
\end{figure}

\subsection{Orbital and charge order}

 The structural responses described above are directly reflected in the orbital and charge degrees of freedom in LCMO. The cooperative JT distortion and its site-selective character can be quantified by introducing an orbital angle at each site $i$, $\Phi_i=\tan^{-1}\!\left(\frac{Q_{2,i}}{Q_{3,i}}\right)$. This orbital angle $\Phi_i$ characterizes the orbital polarization associated with the $Q_2$ and $Q_3$ modes, which lifts the twofold degeneracy of the $e_g$ orbitals and describes the relative contributions of the two JT modes~\cite{Kim2009,Kim2006}.

 Figure~\ref{fig4} shows the evolution of the orbital angle for the two inequivalent Mn sites under various strain conditions. In unstrained LCMO, no pronounced charge or orbital order is found. In this case, the near-degeneracy between the $\ket{x^2-y^2}$ and $\ket{3z^2-r^2}$ states is largely preserved on both sites. Under biaxial strain, however, a cooperative distortion gradually develops, as indicated by the symmetric evolution of the two sites toward $\ket{3y^2-r^2}$ and $\ket{3x^2-r^2}$ character, respectively (see the black curve in Fig.~\ref{fig4}). This corresponds to the development of a staggered orbital order.

 For uniaxial strain, we again find a strong directional dependence. Under strain along (100), similarly to the biaxial case, cooperative JT distortion progressively develops in a staggered pattern. In particular, above 4\% strain, a clear tendency toward the $\ket{3y^2-r^2}$/$\ket{3x^2-r^2}$ orbital order emerges. By contrast, under strain along (010), no such staggered orbital pattern develops; instead, both sites evolve toward a more pronounced $\ket{x^2-y^2}$ character. This asymmetric response to in-plane strain originates from the orthorhombic symmetry of the system, consistent with the structural response discussed above. Strain applied along (100) is primarily accommodated within the $ab$ plane, enhancing the in-plane octahedral rotation and thereby promoting staggered orbital order. In contrast, strain along (010) is absorbed more strongly through out-of-plane tilting, as shown in Fig.~\ref{fig3}(f), which suppresses the tendency toward staggered in-plane orbital order. This contrast is also evident in the partial charge densities shown in Fig.~\ref{fig5}: under strain along (100), the $\ket{3y^2-r^2}$/$\ket{3x^2-r^2}$ orbital order is clearly developed, whereas under strain along (010), the $\ket{x^2-y^2}$ character is preferred.
 
 For strain along the (110) and ($\bar1\bar10$) directions, yet another type of orbital order emerges. As discussed above, the two inequivalent Mn sites become strongly differentiated: one site develops orbital polarization toward $\ket{3y^2-r^2}$ or $\ket{3x^2-r^2}$, while the other largely retains $\ket{x^2-y^2}$ character. This establishes new $CF$-type order introduced above, characterized by the coexistence of ferro-orbital order, $C$-type charge disproportionation, and a FM ground state. Such strong site selectivity, already evident in the breathing mode $Q_1$, is also reflected in the asymmetric JT modes $Q_2$ and $Q_3$. The partial charge density for strain along (110) clearly shows the coexistence of $\ket{3x^2-r^2}$-like and $\ket{x^2-y^2}$-like orbital characters in the MnO$_6$ octahedra for enhanced and suppressed $Q_1$ modes, respectively, providing a direct real-space signature of the $CF$-type order (Fig.~\ref{fig5}). This partially polarized orbital state, originating from the asymmetric JT response together with the breathing distortion, may be viewed as intermediate between the states realized under strain along (100) and (010), and gives rise to a distinct direction-selective phase.

\begin{figure}[t]
\centering
\includegraphics[width=\linewidth,clip,trim=1 1 1 1]{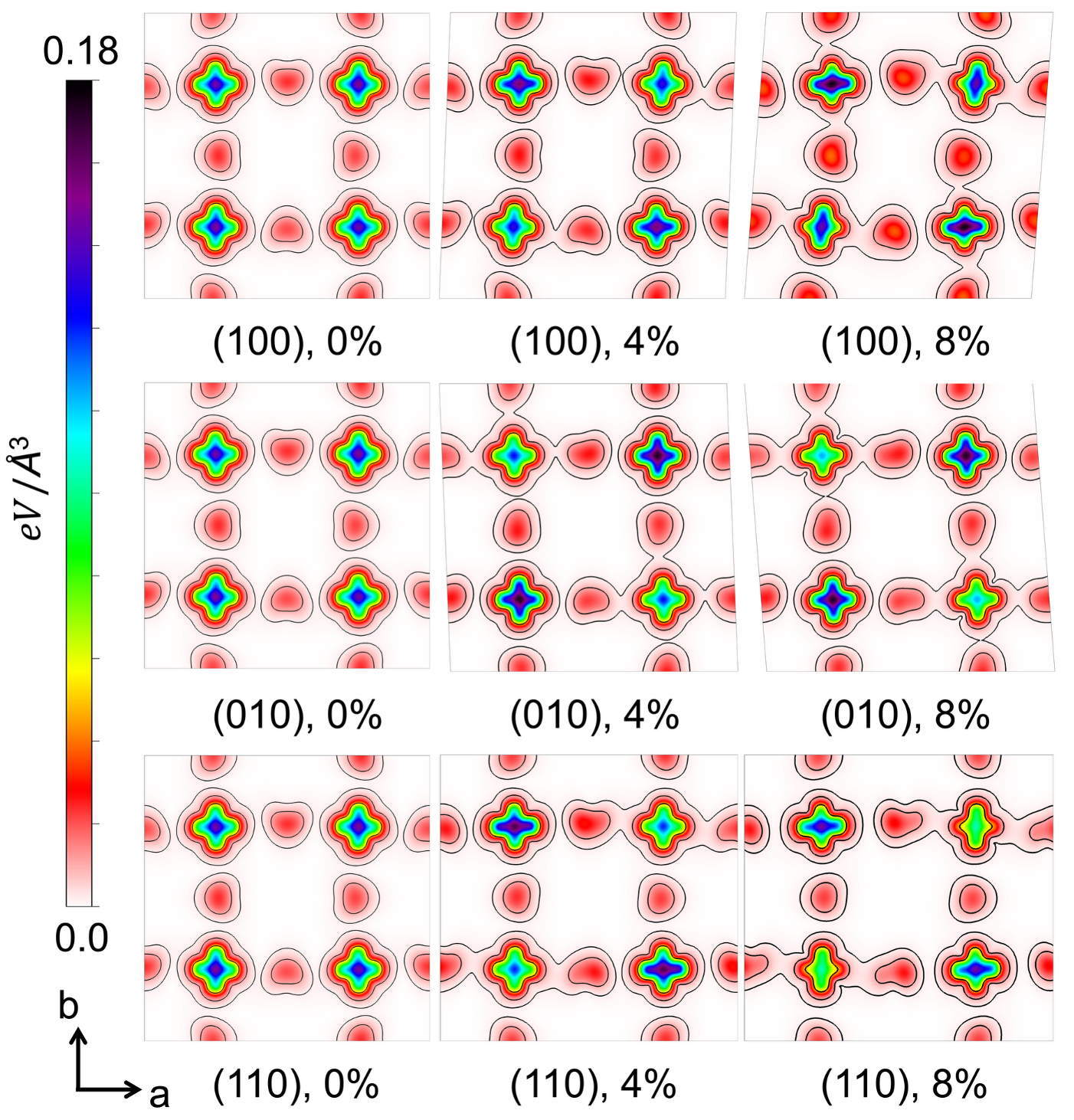}

\caption{\label{fig:left} Partial charge density for the occupied $e_g$ character along the $c$ axis (Fig.~\ref{fig2}(b)) is presented. Strain along the (100) and (010) directions stabilizes distinct orbital characters, namely $\ket{3x^2-r^2}$/$\ket{3y^2-r^2}$ and $\ket{x^2-y^2}$, respectively. Further, the two different Mn sites have different electron occupancies. For strain along the (110) and ($\bar1\bar10$) directions, the orbital character is intermediate between those found for the (100) and (010) strain cases.}
\label{fig5}
\end{figure}

 Beyond the orbital shapes themselves, Fig.~\ref{fig5} also reveals the development of charge modulation. For example, under strain along (100), the charge density differs clearly between the two Mn sites, indicating site-selective charge disproportionation. Under strain along (010), where breathing distortion and JT modulation coexist, the charge modulation becomes even stronger. This shows that the inequivalence between the two Mn sites, already identified from the local structural analysis, is also manifested in the charge sector. The difference in the local Mn magnetic moments between the two sites further highlights the direction-dependent nature of the strain-induced charge ordering~\cite{Zheng2003} (see  Supplementary Fig.~S6).

 These unique charge-orbital-lattice patterns are not prominent at strains of 2\% or less. Both orbital polarization and charge modulation become pronounced only beyond this regime. As discussed above, similar threshold behavior is also found in the structural response, where the site-selective breathing mode emerges only at strains of 4\% and higher. This highlights the potential of the extreme-strain technique to access phases beyond the reach of conventional substrate engineering in complex materials.

 Having established the strain-dependent orbital and charge patterns, we now turn to their connection with the magnetic phase transitions.
 
\subsection{Magnetic phase transition}

 Undoped LMO is an $A$-type AFM insulator. This magnetic ground state is commonly understood in terms of Kugel-Khomskii-type superexchange (SE) coupled to cooperative Jahn-Teller distortion and the associated $C$-type orbital order~\cite{Khomskii2021,Pavarini2010}. The staggered $\ket{3x^2-r^2}$/$\ket{3y^2-r^2}$ orbital pattern promotes FM exchange within the $ab$ plane, whereas the ferro-orbital alignment along the $c$ axis favors AFM coupling. Although the tilting and rotation of MnO$_6$ octahedra can delicately modulate the balance of magnetic exchange interactions between neighboring Mn sites~\cite{Kim2011,Jang2018}, the SE mechanism coupled with cooperative JT distortion remains the main contribution to the magnetic ground state.

 Upon hole doping, the partially occupied $e_g$ states become mobile and can hop through the lattice. This delocalization drives the double exchange (DE) mechanism, which stabilizes the emergent FM metallic phase~\cite{Salamon2001,Rivero2016_2}. The robustness of the FM ground state near $x \approx 0.3$ indicates that DE dominates over the competing AFM SE in this doping regime. Under external strain, structural distortions can significantly modify this balance of magnetic exchanges, potentially driving the system away from the FM ground state. This predicted magnetic phase transition is schematically illustrated in Fig.~\ref{fig2}(b).

 Our results show that under biaxial strain, the transition from an FM to an $A$-type AFM state occurs at approximately 4\% strain, accompanied by the emergence of a partially gapped metallic state~(Fig.~S7). In contrast, under uniaxial strain along the (100) and (010) directions, the magnetic transition requires significantly higher strain than in the biaxial case. Remarkably, no phase transition is observed up to 8\% strain when applied along the (110) and ($\bar{1}\bar{1}0$) directions. As shown in Fig.~\ref{fig3}(a) and (c), the magnitudes of both the JT distortion strength ($Q_{\rho,\mathrm{avg}}$) and the average Mn--O bond-length ($Q_{1,\mathrm{avg}}$) are significantly larger under biaxial strain than under uniaxial strain. These enhanced distortions suggest a suppression of $e_g$ orbital hopping, thereby weakening DE, while simultaneously stabilizing the $C$-type orbital order that favors $A$-type AFM SE.

 However, the reason why the FM-to-AFM ground state transition occurs more readily under strain along (100)/(010) directions compared to strain along (110)/($\bar{1}\bar{1}0$) directions cannot be understood solely in terms of the conventional structural parameters $Q_{\rho,\mathrm{avg}}$, $Q_{1,\mathrm{avg}}$, $\phi$, and $\theta$ (Fig.~\ref{fig3}(a), (c), (e), and (f)), which do not show a simple trend across the magnetic phase boundary. Instead, the site-selective components $Q_{\rho,\mathrm{diff}}$ and $Q_{1,\mathrm{diff}}$ show a clearer correlation with the magnetic transition. In particular, smaller site-to-site differences in these modes are generally associated with the stabilization of the AFM phase, whereas larger site selectivity tends to preserve the FM state. This suggests that the $CF$-type charge-orbital reconstruction may serve as an important indicator of the strain-induced magnetic response. However, the structural order parameters themselves do not directly determine the magnetic ground state. Rather, their effect should be understood through the associated modification of orbital-dependent electronic hopping and magnetic interactions.

Because the system lies in a correlated metallic regime with fractional $e_g$ filling, competition between DE from itinerant $e_g$ electrons and SE from multiple $e_g$ and $t_{2g}$ channels, is responsible for the eventual stabilization of the specific magnetic ground state. Within the DE framework discussed by Hotta {\it et al.}~\cite{Hotta2003}, the FM metallic phase is stabilized when the antiferromagnetic SE interaction is sufficiently weaker than the DE interaction mediated by itinerant $e_g$ electrons. The strain-induced FM-to-AFM transition can therefore be understood in terms of a substantial reduction of the DE interaction. In particular, suppression of the itinerant $e_g$ contribution along the $c$ direction naturally favors the $A$-type AFM state, in which the ferromagnetic $ab$ planes are coupled antiferromagnetically along $c$.

 This picture is supported by the evolution of the electronic band structure under strain~[Fig.~S8]. 
 Under 4\% biaxial strain, one of the $e_g$-derived conduction bands crossing the Fermi level along the $\Gamma$--$Z$ path in FM phase shifts upward in energy and the crossing momentum is close to the $\Gamma$ point in contrast with no strain where it is located at the middle of the $\Gamma$--$Z$ path. A similar behavior is observed under (100) and (010) strains, although a larger strain magnitude (8\%) is required. This shift indicates a reduction in the contribution of itinerant $e_g$ carriers associated with motion along the $c$ axis, which is consistent with the emergence of AFM coupling along the $c$ axis.
In contrast, such a reduction of the $e_g$ itinerancy is much less pronounced for strain along the (110) and ($\bar{1}\bar{1}0$) directions, consistent with the persistence of the FM state. These results indicate that the site-selective structural distortions and the associated charge-orbital reconstruction influence magnetism by modifying the orbital-dependent hopping and hence the balance between DE and SE. In this sense, the $CF$-type order itself does not necessarily favor the $A$-type AFM state; rather, the magnetic transition occurs when the accompanying orbital reconstruction sufficiently suppresses the DE contribution, particularly along the $c$ direction.
 
 Finally, we compare our results with the previous study by Hong \textit{et al.}, in which the MIT, inferred from resistivity measurements, was predominantly associated with a transition from the FM to the AFM state~\cite{Hong2020}. In that study, the critical strains were approximately 7\% and 3\% for uniaxial and biaxial strain, respectively, whereas for the $x=0.25$ composition considered here, we obtain somewhat higher critical strains of approximately 9\% and 4\%. This difference can be understood in terms of the doping-dependent competition between the FM metallic and AFM insulating phases as is well-known for hole-doping phase diagram of manganites~\cite{Ramirez1997}. In the lower-doping regime below $x=0.3$, increasing the hole concentration through Ca or Sr substitution strengthens DE and progressively strengthens the FM metallic state up to $x=0.3$ doping. Accordingly, the lower hole concentration in our $x=0.25$ system shifts the strain-induced magnetic transition to a higher strain. At higher doping beyond $x=0.3$, however, competing AFM and charge-orbital-ordered tendencies become increasingly important, thereby reducing the strain required to destabilize the FM metallic state. This trend is also reproduced in our additional calculations for $x=0.375$, as shown in Fig.~S9. Once the AFM insulating phase is established, the band gap increases systematically with increasing strain, as shown in Fig.~S7.

\section{Conclusion}
 In conclusion, we studied La$_{0.75}$Ca$_{0.25}$MnO$_3$ under extreme, direction-selective uniaxial strain and showed that the resulting structural, electronic, and magnetic responses depend strongly on the strain direction. Rather than producing a simple quantitative tuning of a known phase, different uniaxial strain directions stabilize qualitatively distinct intertwined orders through selective control of local octahedral distortions.

 Strain along the (100) and (010) directions leads to contrasting cooperative JT and breathing-dominated responses, while strain along the (110) and ($\bar1\bar10$) directions stabilizes a novel $CF$-type phase with coexisting ferro-orbital order, $C$-type charge disproportionation, and a FM metallic ground state. These results highlight the crucial role of site-selective lattice response in determining the emergent phases.

 Overall, our work establishes extreme uniaxial strain as a highly selective tuning knob for accessing previously unreachable intertwined orders in correlated oxides, and suggests a general route for engineering emergent phases through the coupled charge, orbital, spin, and lattice degrees of freedom.

\begin{acknowledgments}
We thank S. S. Hong for fruitful discussions. We also thank the participants of the conference~[APCTP-2025-C01] held at PCS-IBS, Daejeon, Korea, for fruitful discussions. This work was supported by the National Research Foundation of Korea(NRF) grant funded by the Korea government(MSIT) (Grants No. RS-2024-00401881, No. RS-2026-25472078, No. RS-2023-00256050, and No. RS-2022-NR068223) and KISTI Supercomputing Center (Project No. KSC-2023-CRE-0413). The work at SNU was supported by the Leading Researcher Program of the National Research Foundation of Korea (Grant No. RS-2020-NR049405).

\end{acknowledgments}

\bibliography{main.bib}

\end{document}


\title{Supplementary Information for ``Direction-selective intertwined charge, orbital, and lattice orders under uniaxial strain in hole-doped manganite: La$_{0.75}$Ca$_{0.25}$MnO$_3$"}

\author{Ju Hyeon Lee}
\affiliation{Department of Physics, Kyungpook National University, Daegu 45166, Republic of Korea}
\author{Beom Hyun Kim}
\email{bomisu@gmail.com}
\affiliation{Department of Physics \& Astronomy, Seoul National University, Seoul 08826, Republic of Korea}
\author{Bongjae Kim}
\email{bongjae@knu.ac.kr}
\affiliation{Department of Physics, Kyungpook National University, Daegu 45166, Republic of Korea}
                              

\maketitle

\tableofcontents


\newpage
\section{\label{sec:level1} Structural parameters}

\begin{table*}[h]
\caption{\label{tab:sup1} Comparison of the experimental and calculated structures for different Hubbard parameters. The calculations were performed using the PBEsol functional with a tetragonal $\sqrt{2}\times\sqrt{2}\times2$ unit cell containing four formula unit. In the main text, we used $U=4$~eV and $J_H=1$~eV~\cite{Hong2020,Lee2025}.}
\begin{ruledtabular}
\begin{tabular}{cccc}
Type & $a=b$ ($\mathrm{\AA}$ )& $c$ ($\mathrm{\AA}$) & volume ($\mathrm{\AA^3}$)\\ [0.5ex]
\hline
Exp & 5.459 & 7.701 & 229.49 \\ [0.5ex]
PBEsol & 5.416 & 7.639 & 224.05 \\
PBEsol ($U=4.0$~eV, $J_H=1.0$~eV) & 5.442 & 7.676 & 227.32 \\
PBEsol ($U=4.0$~eV, $J_H=0.7$~eV) & 5.436 & 7.668 & 226.60 \\
PBEsol ($U=6.0$~eV, $J_H=1.0$~eV) & 5.455 & 7.695 & 228.95 \\ [0.5ex]
\end{tabular}
\end{ruledtabular}
\end{table*}

\newpage
\section{\label{sec:level1} Strain configurations}
\begin{figure}[h]
\centering
\includegraphics[width=0.6\linewidth,clip,trim=2 2 2 2]{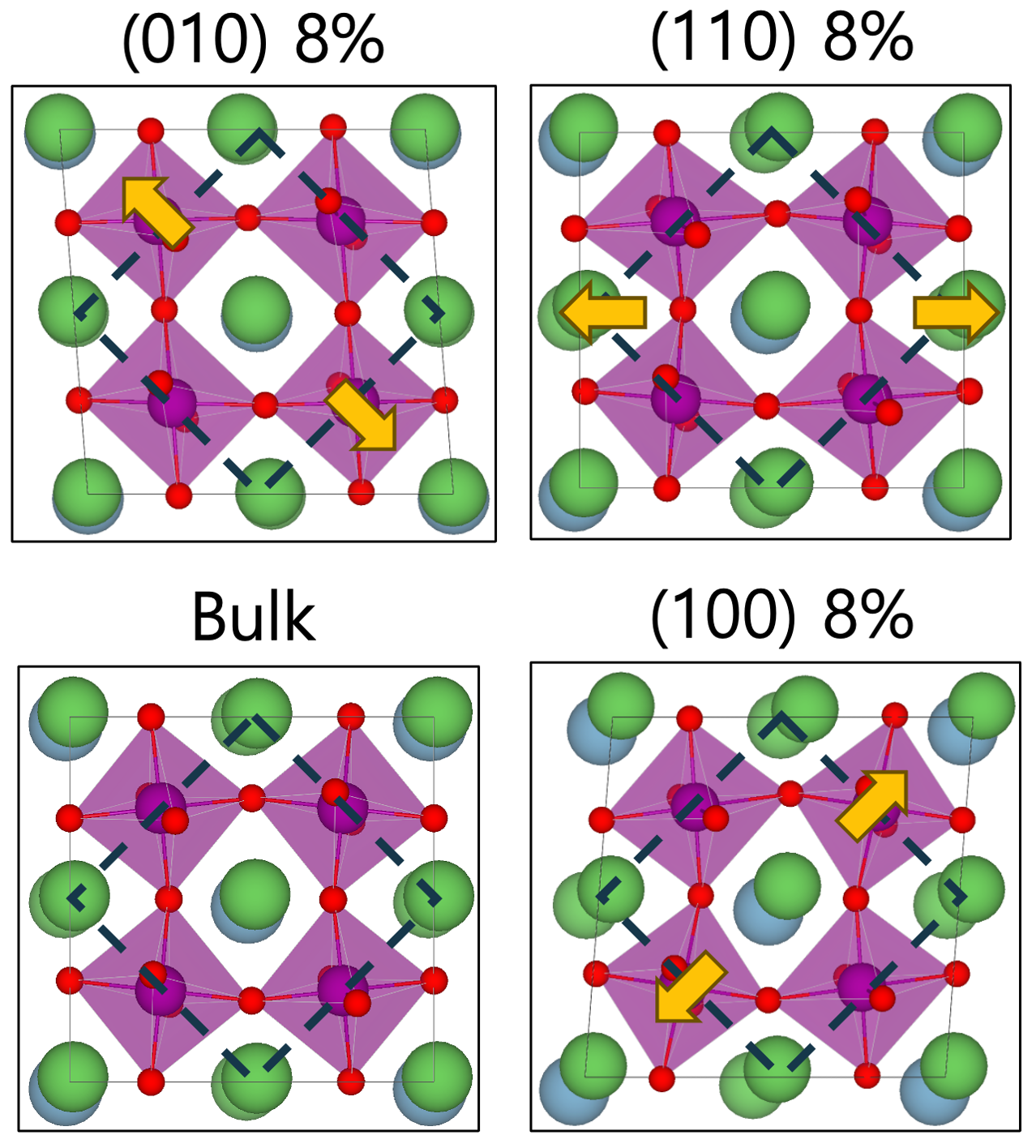}
\caption{The $\sqrt{2}\times\sqrt{2}\times2$ orthorhombic unit cell is outlined by the black dashed rectangle. The corresponding pseudocubic supercell and the applied strain directions are indicated by the black line and yellow arrows, illustrating the geometric relation between the orthorhombic and pseudocubic coordinate systems.
}
\label{fig: Sup1} 
\end{figure}

\newpage
\section{\label{sec:level2} Relative energy of magnetic states}
\begin{figure}[h]
\centering
\includegraphics[width=0.6\linewidth,clip,trim=1 1 1 1]{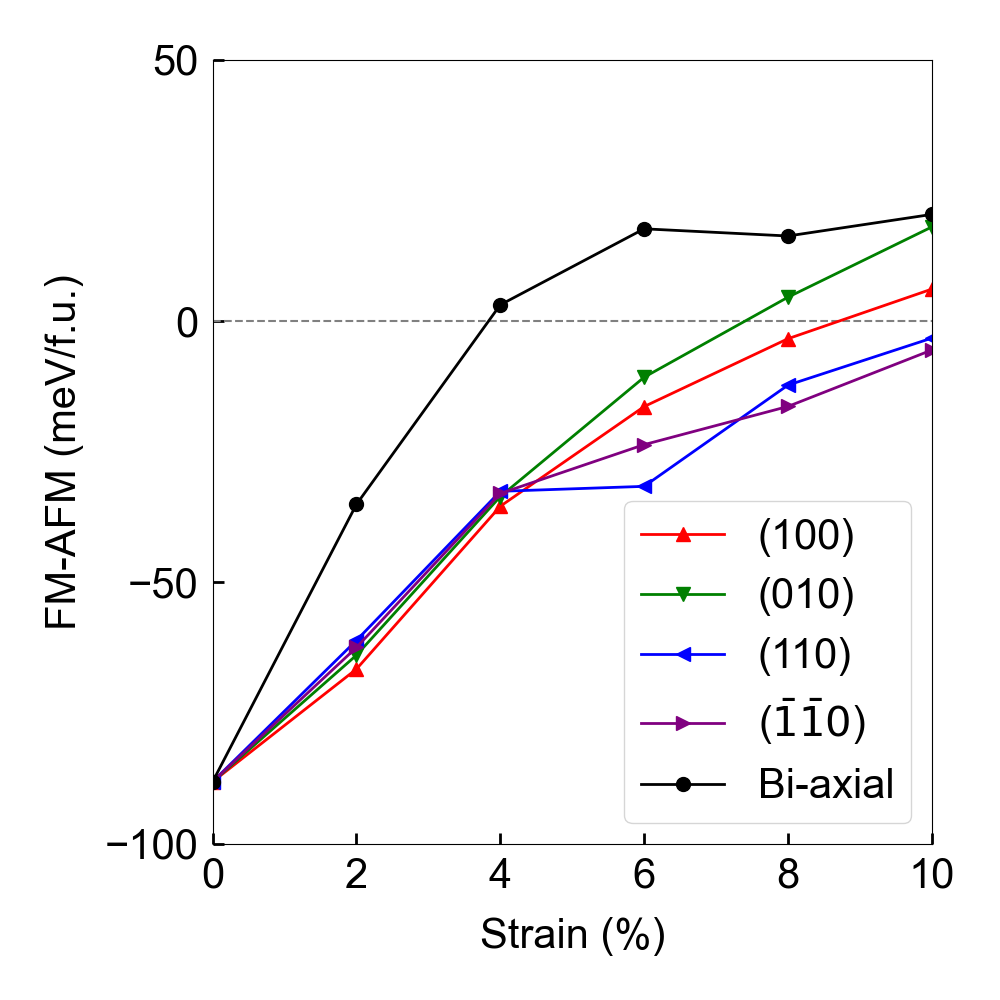}
\caption{Relative energy difference between the FM and $A$-type AFM states for La$_{0.75}$Ca$_{0.25}$MnO$_3$. The black, red, green, blue, and purple curves denote biaxial strain and uniaxial strain applied along the (100), (010), (110), and ($\bar1\bar10$) directions, respectively.}
\label{fig: Sup2} 
\end{figure}

\newpage
\section{\label{sec:level3}Lattice parameters}
\begin{figure}[h]
\centering
\includegraphics[width=\linewidth,clip,trim=1 1 1 1]{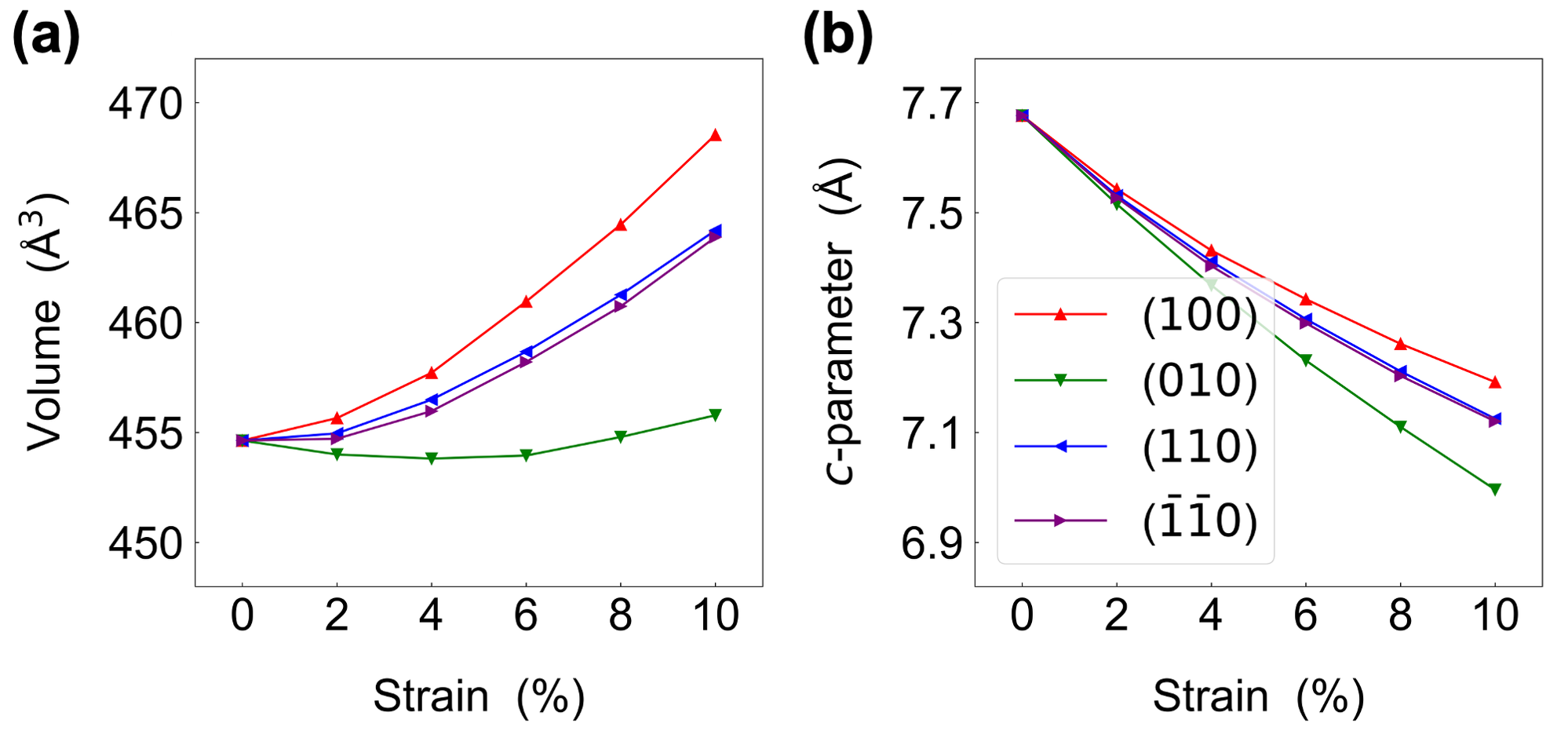}
\caption{Strain dependence of the volume and lattice parameter c, shown in (a) and (b), respectively. The red, green, blue, and purple curves correspond to uniaxial strain along the (100), (010), (110), and ($\bar1\bar10$) directions, respectively.}
\label{fig: Sup3}
\end{figure}

\newpage
\section{\label{sec:level4} Mn--O Bond lengths}

\begin{figure}[h]
\centering
\includegraphics[width=0.35\linewidth,clip,trim=1 1 1 1]{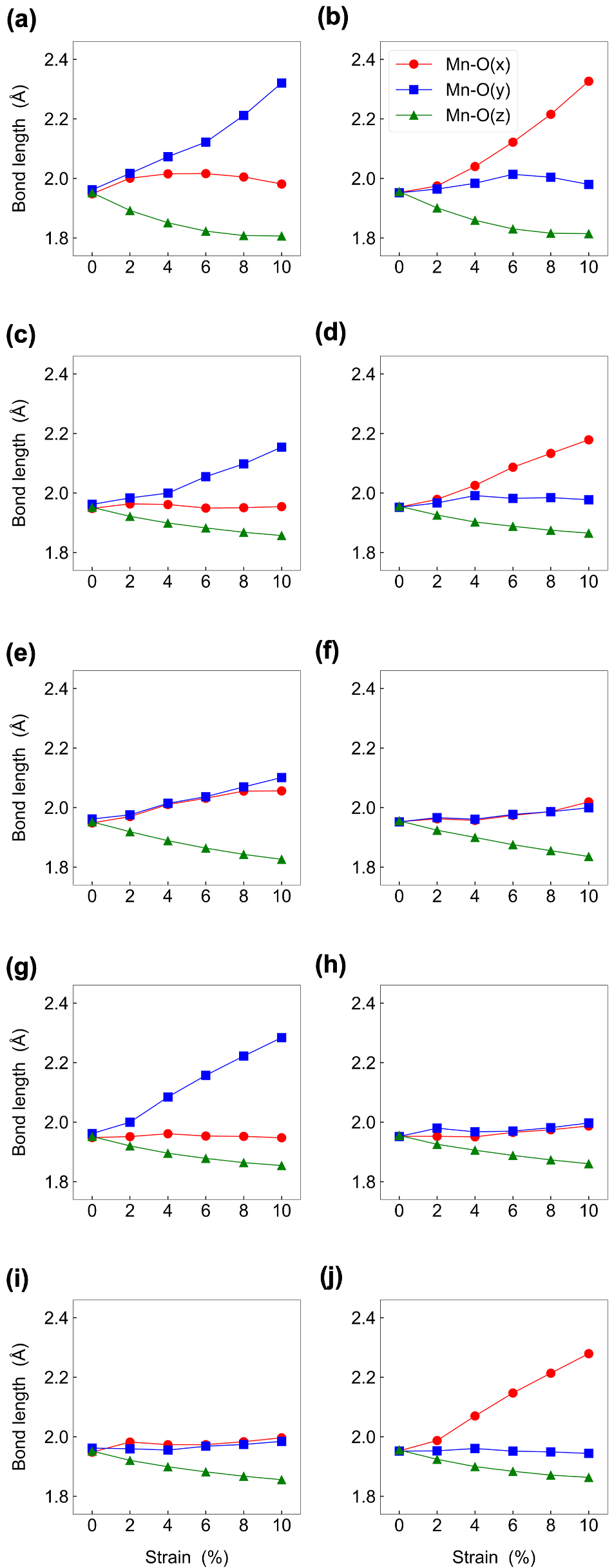}
\caption{Mn--O bond lengths as functions of strain. Panels (a), (c), (e), (g), and (i) correspond to octahedral site 1, while panels (b), (d), (f), (h), and (j) correspond to site 2. Panels (a) and (b) show the biaxial-strain case; (c) and (d), strain along (100); (e) and (f), strain along (010); (g) and (h), strain along (110); and (i) and (j), strain along ($\bar1\bar10$). The red, blue, and green curves denote the Mn--O bond lengths $X$, $Y$, and $Z$ respectively, along the local axes $x$, $y$, and $z$ (See Fig.~2 (a)).}
\label{fig: Sup4} 
\end{figure}

\newpage
\section{\label{sec:level5} Structural distortion and Mn--O--Mn angles in LaMnO$_3$}
\begin{figure}[!h]
\centering
\includegraphics[width=0.5\linewidth,clip,trim=1 1 1 1]{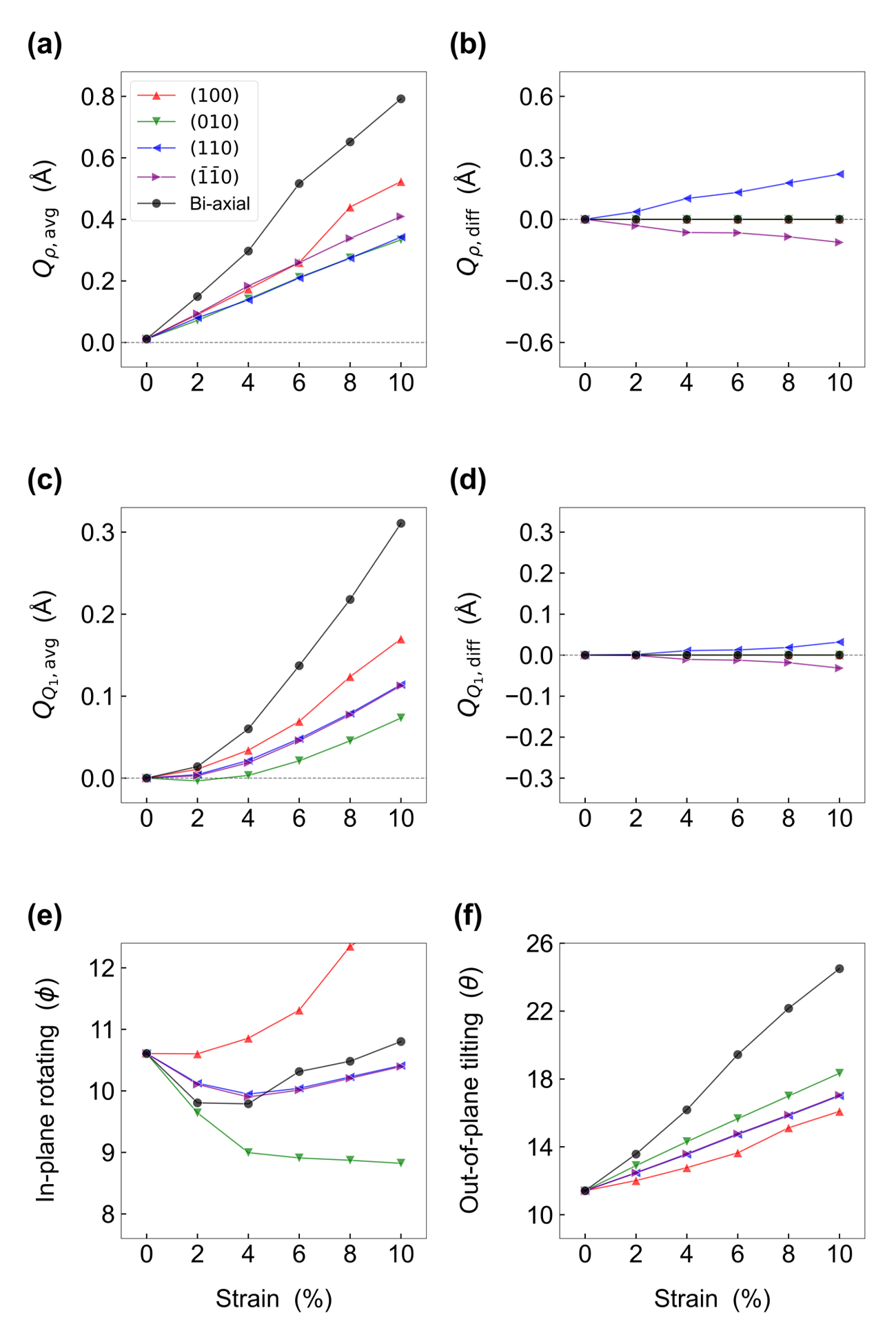}
\caption{Jahn-Teller distortion strengths are shown in (a) and (b), corresponding to the average and site-difference components, respectively. Panels (c) and (d) show the $Q_1$ distortion for the two octahedral sites, while (e) and (f) show the Mn--O--Mn angles associated with the in-plane rotation ($\phi$) and out-of-plane tilt ($\theta$), respectively. The corresponding angles and octahedral sites are illustrated in Fig.~2(a). For the undoped case, the Ca atoms in the LCMO cell were replaced with La, and the structure was fully relaxed under the same constraints on the $a$ and $b$ lattice parameters as described in the main text. The black, red, green, blue, and purple curves denote biaxial strain and uniaxial strain applied along the (100), (010), (110), and ($\bar1\bar10$) directions, respectively.}
\label{fig: Sup5}
\end{figure}

\newpage
\section{\label{sec:level6} Magnetic moments of strained La$_{0.75}$Ca$_{0.25}$MnO$_3$}

\begin{figure}[h]
\centering
\includegraphics[width=\linewidth,clip,trim=1 2 2 2]{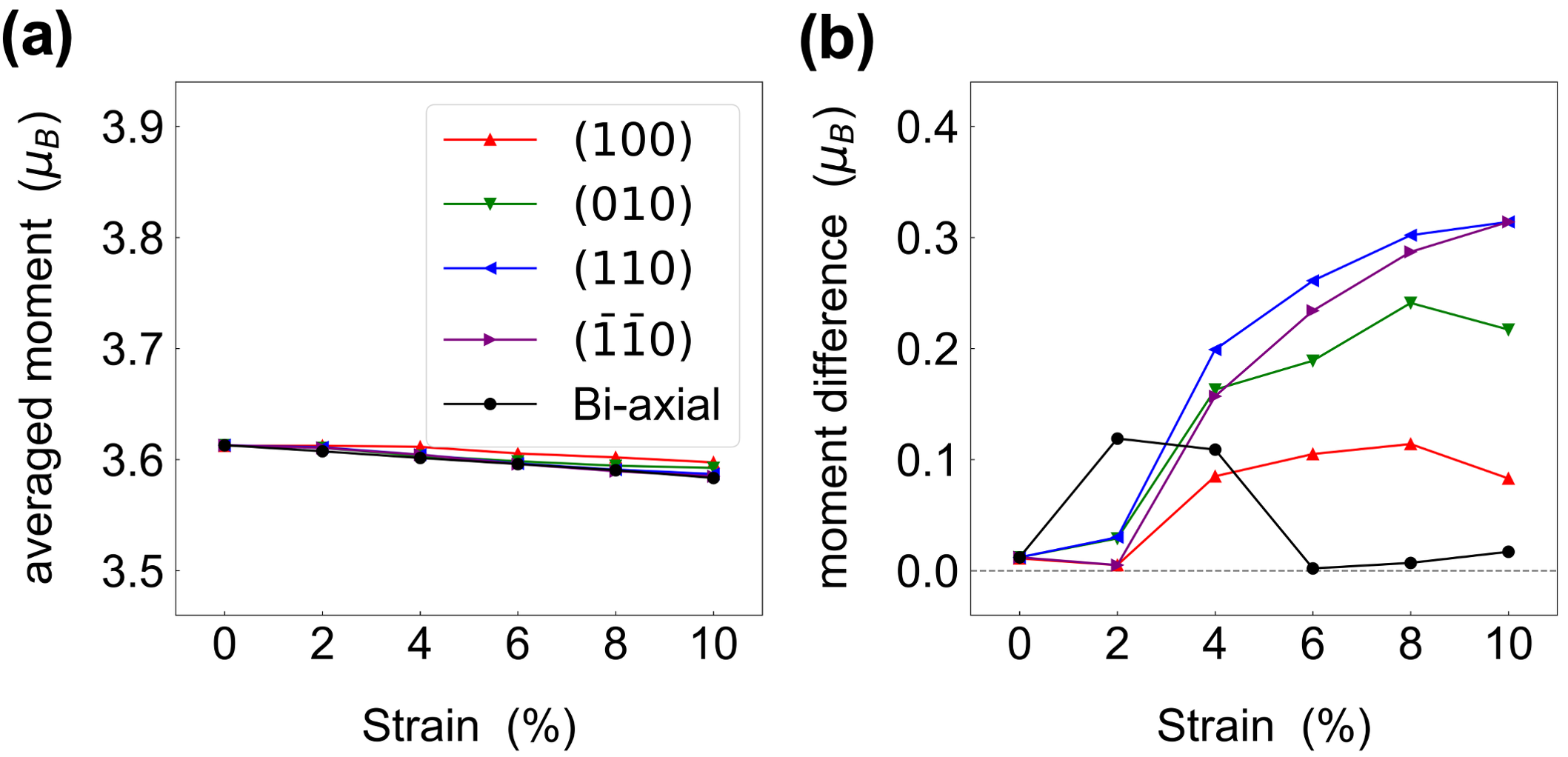}
\caption{The average magnetic moments and their site differences are shown in (a) and (b), respectively. The black, red, green, blue, and purple curves denote biaxial strain and uniaxial strain applied along the (100), (010), (110), and ($\bar1\bar10$) directions, respectively.}
\label{fig: Sup6}
\end{figure}

\newpage
\section{\label{sec:level7} Electronic band structures}

\begin{figure}[h]
\centering
\includegraphics[width=0.9\linewidth,clip,trim=2 1 1 2]{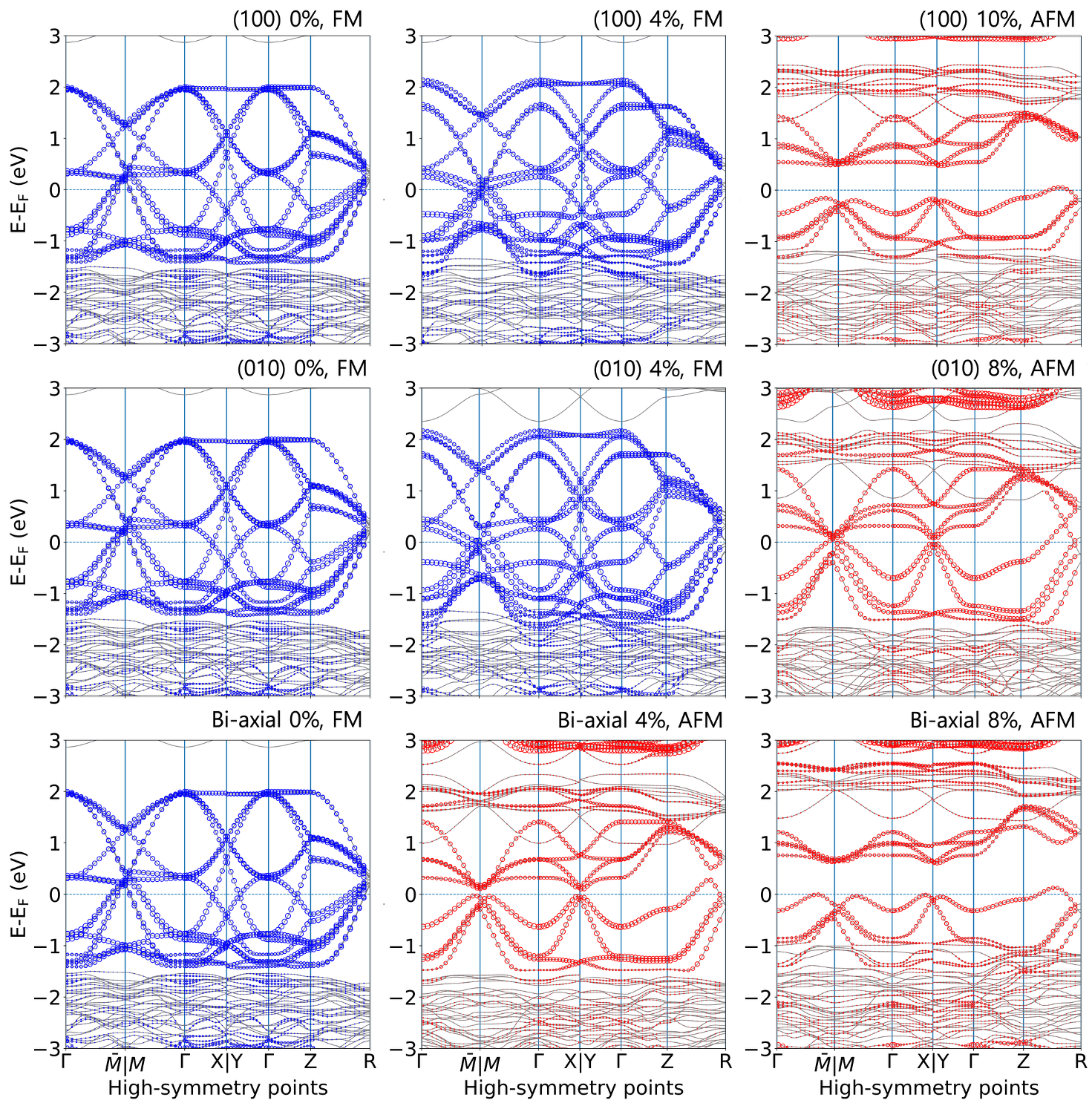}
\caption{ The band structure along high-symmetry lines is presented. The $X$, $Y$, $M$, $\bar M$, and $R$ are symmetrically relevant high-symmetry points. Upon the magnetic transition, a partially gapped state develops predominantly near the $\bar{M}|M$ and $X|Y$ points for the (100) and biaxial strain cases. In contrast, for strain along (010), no gap opens near $\bar{M}|M$, and the system remains metallic. The blue and red circles denote $e_g$ orbital characters at FM and A-AFM phases, respectively.}
\label{fig: Sup7}
\end{figure}

\newpage
\section{\label{sec:level8} Band structure for FM and A-AFM phases}

\begin{figure}[h]
\centering
\includegraphics[width=0.8\linewidth,clip,trim=2 0 1 1]{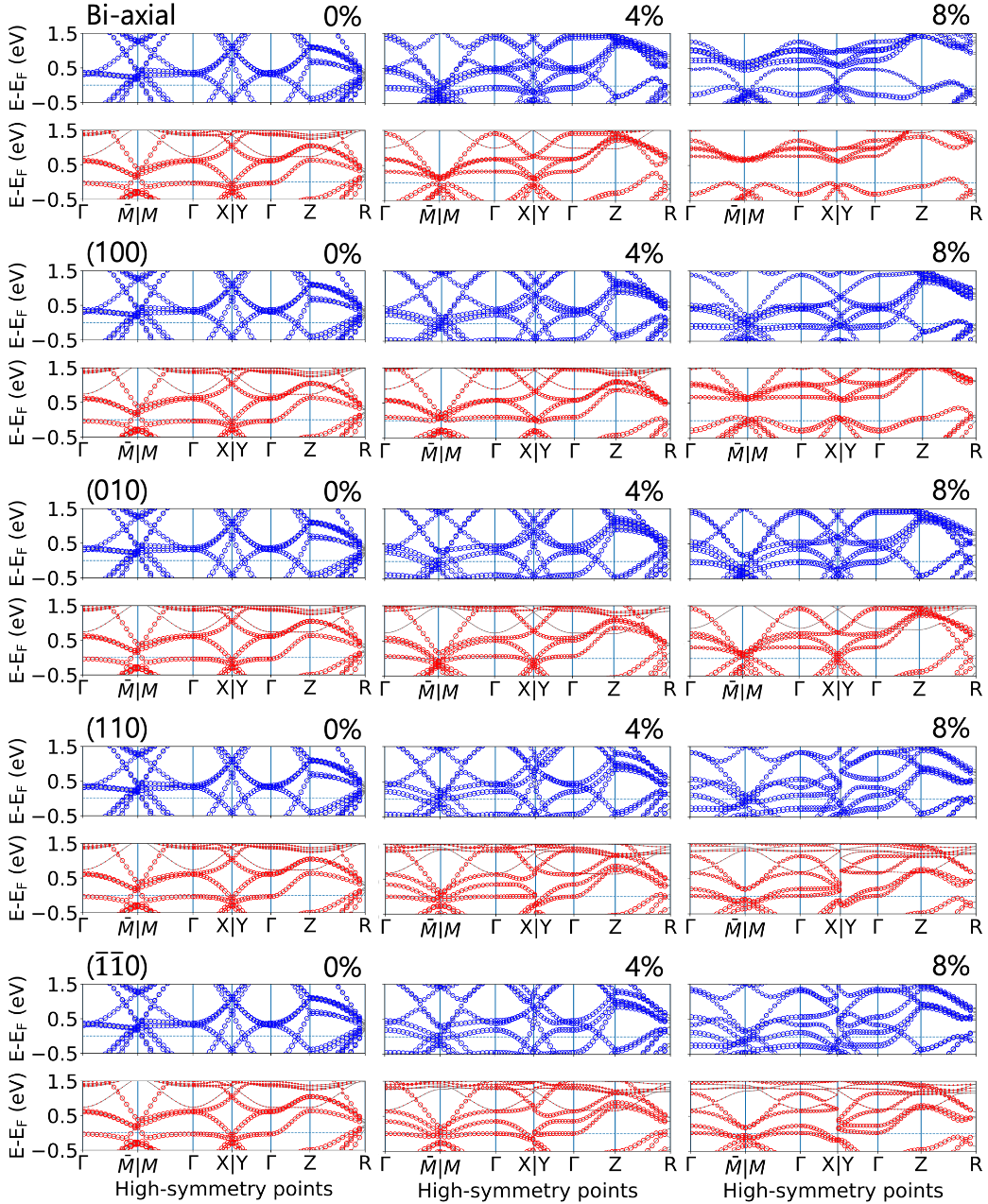}
\caption{Electronic band structures under various strain directions are presented. The blue and red circles denote the $e_g$ orbital characters in the FM and $A$-type AFM phases, respectively. With increasing strain, the FM $e_g$-derived band crossing the Fermi level along the $\Gamma$-$Z$ path becomes less dispersive, indicating a reduction of the corresponding $c$-axis itinerancy.}
\label{fig: Sup8}
\end{figure}

\newpage
\section{\label{sec:level9} Relative energy of magnetic phases in La$_{0.625}$Ca$_{0.375}$MnO$_3$}

\begin{figure}[h]
\centering
\includegraphics[width=\linewidth,clip,trim=1 1 1 1]{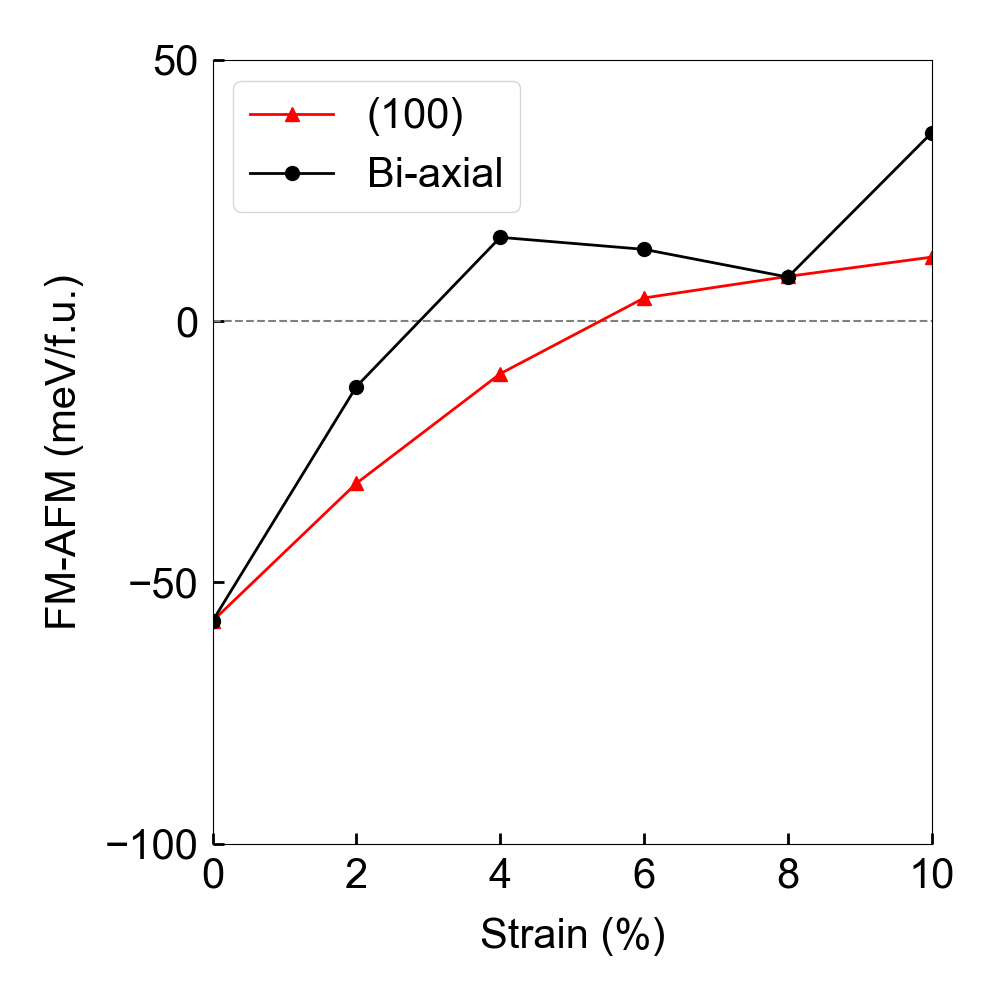}
\caption{Relative energy difference between the FM and $A$-type AFM states for La$_{0.625}$Ca$_{0.375}$MnO$_3$. The black, red curves denote biaxial strain and uniaxial strain applied along the (100) direction.}
\label{fig: Sup9}
\end{figure}

\bibliography{sup}